\documentclass[prl,twocolumn,letterpaper]{revtex4}
\usepackage{epsfig,amssymb,amsfonts}
\newcommand{\ket}[1]{\left | \, #1 \right\rangle}
\newcommand{\bra}[1]{\left \langle #1 \, \right |}
\newcommand{\proj}[1]{\ket{#1}\!\bra{#1}}

\begin{document}
\title{Cluster state quantum computation in coupled cavity arrays}
\author{Dimitris G. \surname{Angelakis}}
    \email{dimitris.angelakis@qubit.org}
    \affiliation{Centre for Quantum Computation,
             DAMTP,
             Centre for Mathematical Sciences,
             University of Cambridge,
             Wilberforce Road,
             Cambridge CB3 0WA, UK}
\author{Alastair \surname{Kay}}
\affiliation{Centre for Quantum Computation,
             DAMTP,
             Centre for Mathematical Sciences,
             University of Cambridge,
             Wilberforce Road,
             Cambridge CB3 0WA, UK}

\begin{abstract}
Measurement-based quantum computation has revolutionized quantum
information processing, and the physical systems with which it can
be implemented. One simply needs the ability to prepare a particular
state, known as the cluster state, and subsequently to perform
single-qubit measurements on it. Nevertheless, a scalable
implementation is yet to be realized. Here we propose a hybrid
light-matter system comprised of coupled cavities interacting with
two level systems. Utilising the stable, individually addressable,
qubits resulting from the localised long-lived atom-photon
excitations, we demonstrate how to use the natural system dynamics
to `weave' these qubits into a cluster state and propose the
implementation of quantum algorithms employing just two rows of
qubits. Finally, we briefly discuss the prospects for experimental
implementation using atoms, quantum dots or Cooper pair boxes.
\end{abstract}
\maketitle

Quantum computation on cluster states \cite{cluster1} has been
proposed in a variety of systems, including linear optics, quantum
dots, neutral atoms in optical lattices, and flying atom schemes
\cite{nielsen:040503,browne-2005-95,kok:04,lim-2005-95,benjamin-2005-7,
borhani-2005-71,Kay:2005a,blythe-varcoe,schon}. To date, experiments
have been performed using optical lattices \cite{Greiner:2003a},
where the cluster state can be created, but the current lack of
individual addressing remains the stumbling block and linear optics
\cite{Walther,Pan}, where scalability remains a problem due to the
need to generate the initial many-photon state from, for example, high orders of the parametric down conversion proecss. On the other hand,
there have recently been theoretical and experimental breakthroughs
into the possibility of direct coupling of high Q cavities and in
achieving strong coupling between the cavity mode and an embedded
two-level system. A variety of technologies have been employed,
namely fiber coupled micro-toroidal cavities interacting with atoms
\cite{aoki-2006,armani-vahala03}, arrays of defects in photonic band
gap materials (PBGs)
\cite{vuckovic-apl,song-noda05,AntonioBadolato05202005} and
superconducting qubits coupled through microwave stripline
resonators \cite{wallraff-2004}. This has prompted proposals for the
implementation of optical quantum computing
\cite{angelakis-ekert04}, the production of entangled photons
\cite{angelakis-bose06} and the realization of Mott insulating and
superfluid phases
\cite{angelakis-bose06b,hartmann-2006-2,greentree-2006}. Here we
propose the use of such arrays for the realization of cluster state
quantum computation.
\begin{figure*}
\includegraphics[width=0.98\textwidth]{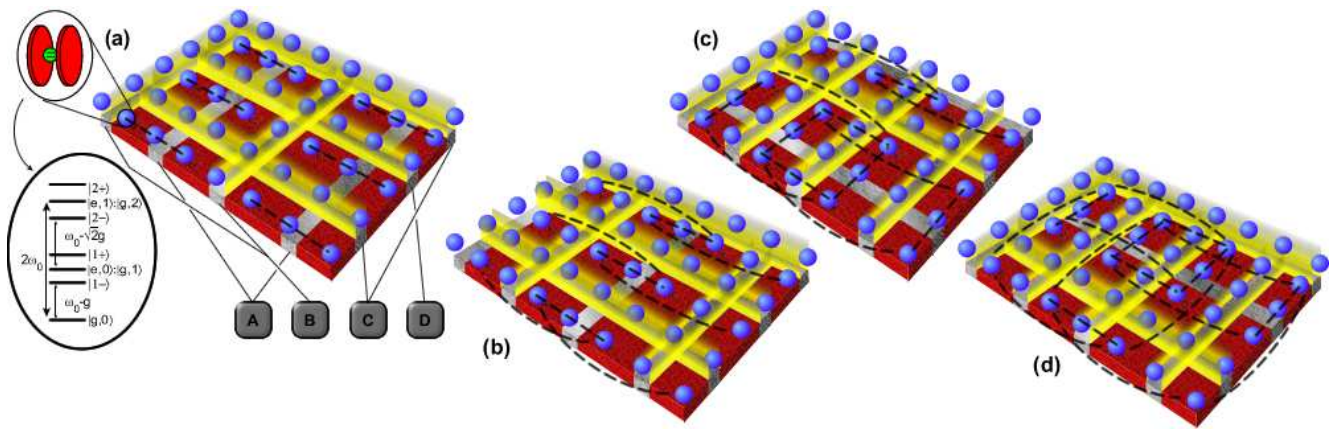}
\caption{We work with a 2D array of atom-cavity systems. When the
atom is on resonance with the cavity, the ground state $|g,0\rangle$
and the first excited state $|1-\rangle$ of the combined atom-photon
(polaritonic) system in each site can be used as qubits. By applying
Stark shifts with control electrodes or properly tuned laser fields
to sets of qubits (the gates shown under the qubits), we disable the
exchange Hamiltonian of a qubit to all of its neighbours. Applying
gates B, C, D or equally switching off gate A(part (a)) creates
chains of 3 qubits which apply controlled-phases and SWAPs between
the qubits at either end of the chain (indicated by dashed lines).
Repeating the sequence with the other 3 gates  is sufficient to
connect the 3-chains and generate a cluster state in parallel across
the whole device. Single qubit rotations and measurements are made
by properly applying local external fields, utilizing the fact that
the cavities can be well separated. } \label{pol} \vspace{-0.5cm}
\end{figure*}

{\em System Description: } We start be describing the system and
showing how to construct qubits from the hybrid light-matter
excitations (polaritons). For simplicity, we describe the system as
a linear chain of $N$ coupled cavities doped with two level systems,
although this is readily adapted to the two-dimensional setting that
we require. $\ket{g}_{k}$ and $\ket{e}_{k}$ are the atomic ground
and excited states at site $k$ (we henceforth use the term `atom' to
refer to any relevant two level system).  The Hamiltonian describing
the system is the sum of three terms; $H^{free}$ is the Hamiltonian
for the free light and dopant parts, $H^{int}$ the Hamiltonian
describing the internal coupling of the photon and dopant in a
specific cavity and $H^{hop}$ for the light hopping between
cavities.
\begin{eqnarray}
H^{free}&=&\omega_{d}\sum_{k=1}^N a_k^\dagger a_k+\omega_{0}\sum_k\proj{e}_{k} \\
H^{int}&=&g\sum_{k=1}^N(a_k^\dagger\ket{g}_{k}\bra{e}_{k}+a_k\ket{e}_k\bra{g}_k)\\
H^{hop}&=&A\sum_{k=1}^N(a_k^\dagger a_{k+1} +a_k a_{k+1}^\dagger)
\end{eqnarray}
$\omega_{d}$ and $A$ are the photon frequencies and hopping rates
respectively and g is the light-atom coupling strength. The
$H^{free}+H^{int}$ component of the Hamiltonian can be diagonalized
in a basis of combined photonic and atomic excitations, called {\it
polaritons} (Fig.~\ref{pol}). These polaritons are defined by
creation operators $P_{k}^{(\pm,n)\dagger}=\ket{n\pm}_k\bra{g,0}_k$,
where the polaritons of the $k$th atom-cavity system are given by
$\ket{n\pm}_k=(\ket{g,n}_k\pm \ket{e,n-1}_k)/\sqrt2$ with energies
$E^{\pm}_{n}=n\omega_{d}\pm g\sqrt{n}$, and $\ket{n}_k$ denotes  the
$n$-photon Fock state. As has been shown elsewhere, a polaritonic
Mott phase exists in this system where a maximum of one excitation
per site is allowed \cite{angelakis-bose06b}. This originates from
the repulsion due to the photon blockade effect
\cite{PhysRevLett.79.1467,birnbaum-2005}. In this Mott phase, the
system's Hamiltonian can be written in the interaction picture as $
H_{I}=A\sum_{k=1}^{N-1}P^{\dagger}_{k}P_{k+1}+P_{k}P^{\dagger}_{k+1},$
where $P_{k}^{\dagger}=P_{k}^{(-,1)\dagger}$ (Fig.~\ref{pol}). As
double or more occupancy of the sites is prohibited, one can
identify $P^{\dagger}_{k}$ with
$\sigma^{+}_k=\sigma^x_k+i\sigma^y_k$, where $\sigma^x_k$ and
$\sigma^y_k$ are the standard Pauli operators. The system's
Hamiltonian then becomes the standard XY model of interacting spin
qubits with spin up/down corresponding to the presence/absence of a
polariton.
\begin{eqnarray}
H_I=A\sum_{k=1}^{N-1}\sigma^x_k\sigma^x_{k+1}+\sigma^y_k\sigma^y_{k+1}.
\end{eqnarray}
Some applications of XY spin chains in quantum information
processing can thus been implemented in this system \cite{bose}.

{\em Cluster state generation: } The typical implementation of
cluster state quantum computing requires initializing all qubits in
a 2D lattice in the $\ket{+}=(\ket{0}+\ket{1})/\sqrt{2}$ state and
then performing controlled-phase gates ($CP$) between all
nearest-neighbours. In the present system, we have no direct
two-qubit gate and the available interaction is not of the Ising
type, which straightforwardly gives controlled-phase gates
\cite{cluster1}, but an `always on' global Hamiltonian coupling of
the XY form.  It behooves us to consider how we will
proceed with the measurement sequence once the cluster state has
been generated without the system continuing to evolve. (The need to
disable the evolution is an aspect often neglected when discussing
the generation of cluster states, either as the ground states of
`natural' Hamiltonians \cite{rudolph:06}, or from evolution of the
Ising Hamiltonian). This requirement can be realized by combining
the system's natural dynamics with a protocol where some of the
available physical qubits are used as gate ``mediators"  and the
rest as the logical qubits. The mediator atoms can be Stark shifted on
and off resonance from their cavities through
the application of an external field, inhibiting the photon
hopping and thereby isolating each logical qubit. The same inhibition
of couplings will be used to generate the cluster state. We note
here that the error introduced in the step is due to a second-order
transition between on-resonance qubits (via a dark-passage through
the central off-resonant qubit), which is thus supressed by a factor of
order $A/\Delta$, where $\Delta$ is the detuning of the off-resonant
cavity.

Before describing the 4-step global gate sequence to create the
cluster state, first observe that for the control phase part is
enough to localize chains of 3 qubits, let them evolve for a time
$t_0=\pi/(2\sqrt{2}A)$ and then apply a measurement on the middle
`mediator' qubit (in the $\sigma^z$ basis). Depending on the measurement result, $\ket{0}$ or $\ket{1}$, a nonlocal gate
is generated between the remaining two qubits, either {\sc SWAP}.$(\sigma^z\otimes\sigma^z).CP$ or {\sc
SWAP}.$CP$ respectively\cite{yung-2004-4,Christandl:2004a}. In both
cases, the gates in addition to the $CP$ are Clifford operations
which can be recorded and taken into account during the
measurement-based computation.

Our sequence to generate the cluster state initiates by preparing
all qubits in the $\ket{+}$ state through the application of global
$\pi/2$ pulse. One quarter of the sites will be used as logical
qubits and the rest as ``mediators" and ``off" qubits
interchangeably.   All qubits addressed by the gates A to D
(Fig.~\ref{pol}) are, by default, ``off'', thereby isolating all the
qubits. Switching on any one of the four gates thus creates chains
of 3 qubits, which we use to enact a $CP$ between pairs of qubits
(separated by a mediator qubit, which was previously off).
Consecutive use of each of the gates A to D serves to enact a $CP$
gate between a particular qubit and all of its nearest-neighbours,
and this happens in parallel across the whole device.
This entire sequence is illustrated in Supplementary Video 1. The
measurement sequence is then applied as requested by the cluster
state algorithm, utilizing the local accessibility of the sites (in any implementation, the cavity-atom systems are
well separated compared to the resolution of the external field used
for addressing
them) \cite{aoki-2006,armani-vahala03,vuckovic-apl,song-noda05,
AntonioBadolato05202005,wallraff-2004}.

\begin{figure}
\includegraphics[width=0.4\textwidth]{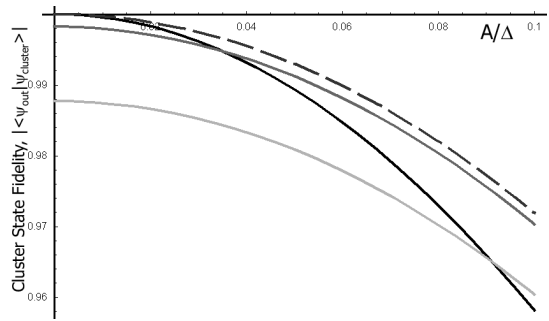}
\caption{The fidelity of generation of a cluster state on a 3x3 grid
of cavities, as the detuning $\Delta$ of the mediator off-resonance
cavities is varied(in units of the hopping A). The dashed line
includes post-selection on getting $\ket{0}$ outcomes when measuring
off-resonance qubits. The gray lines also incorporate spontaneous
decay and cavity leakage of $0.05A$ (dark) and $0.08A$ (light).}
\label{simulation} \vspace{-0.5cm}
\end{figure}

In Fig.~\ref{simulation}, we calculate the fidelity of generation of
a cluster state on a 3x3 array of cavities. More sophisticated
schemes have the potential to further reduce the experimental
errors. For example, standard Hamiltonian simulation techniques
allow us to negate the second order exchange term due to the
off-resonance cavities, simply by repeatedly applying $\sigma_z$ gates
to every second on-resonance triplet throughout the evolution. One
might even hope that we could use this coherent effect to enhance
the scheme through the use of, for example, optimal control
techniques.
Most of the errors considered here (cavity leakage, spontaneous
emission of the atom, and on-off detuning of qubits) are local
effects, introducing local noise, which can ultimately be addressed
by fault-tolerant techniques \cite{raussendorf}.

{\em Implementing algorithms: } Initial experimental algorithmic
implementations with coupled cavities can be expected to utilize the
most basic building block of our scheme, a $3\times 3$ grid of
cavities, which allows us to generate a four-qubit cluster state. As
with the four-photon cluster state recently used by Walther
{\em et al.} \cite{Walther}, this cluster state would be suitable
for demonstrating the preparation of an arbitrary one-qubit state, an entangling gate between two qubits,
and even the implementation of Grover's search algorithm on
two qubits \cite{Walther}. For example, by applying the local gates
$H\otimes H\otimes\sigma_z\otimes\sigma_z$, where $H$ is the Hadamard rotation,
we convert of `box' cluster that the $3\times 3$ grid prepares into the 1D
cluster state of 4 qubits, which is given the interpretation of a
single qubit, and measurements on the state yield quantum gates on
this single qubit. Moreover, generation of this four qubit cluster
state is simpler than generation of an arbitrarily sized cluster
state because we only need two control steps instead of four,
thereby keeping us even further within the decoherence time of the
system.

\begin{figure}
\includegraphics[width=0.3\textwidth]{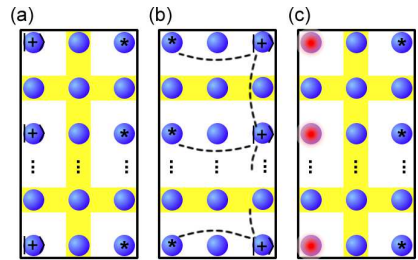}
\caption{ Sequence for minimising the number of qubits required for
a cluster state computation. (a) After the first $n-1$ steps of the
algorithm, the first column of qubits is initialised in the
$\ket{+}$ state, and the third column, with qubits denoted by $*$,
are in the state of output for the first $n-1$ steps of the
computation. (b) We use control sequences, bringing mediator qubits
on resonance, to convert the $\ket{+}$ states into a cluster state,
and to entangle them with the output qubits. The SWAP in the
entangling operation moves these output qubits to the first column.
(c) Measure the qubits of the first column as corresponds to the
$n^{th}$ step of the computation, and reinitialise in the $\ket{+}$
state. The rightmost column corresponds to the output. The sequence
then repeats.} \label{panels} \vspace{-0.5cm}
\end{figure}

Perhaps the next important step would then be to demonstrate Shor's
factoring algorithm, the factoring of 15 being the standard
demonstration. To implement as a cluster state computation, the six
computational qubits \cite{shor_imp} translate into the requirement
of a cluster state that is eleven qubits wide. Hence, we need an
array which is 21 cavities wide. The breadth of the cluster state,
which corresponds to time in the circuit model, is a quantity that
we can trade against the time taken for the computation. At one
extreme, we can create the whole cluster state in one go, with the
simple set of four steps already outlined, and we benefit from the
large degree of parallelism available to us. This requires a 2D grid
of cavities of size $21\times 311$ \footnote{To arrive at this
required number of gates, we have taken the circuit presented in
\cite{shor_imp} and converted it into a nearest-neighbor, 2-qubit
gate algorithm. Hence, the possibility for some small degree of optimisation in the number of qubits remains.}. At the other extreme, a grid of
$21\times 3$ cavities suffices. In this case, one starts with the
$11\times 2$ cluster state, and performs one time step of
measurement (i.e.~measure the 11 qubits in one column). The result
remains in the other column. We then repeat the cluster state
generation process, reinitialising the measured qubits in the
cluster state, and performing the next time step
(Fig.~\ref{panels}). This requires 156 consecutive entangling steps,
but the reinitialising of the cluster state after measurement
eliminates the effect of decoherence over this timescale. Any
combination between these two extremes is also possible, and is a
necessary property of any scalable implementation of cluster state
computation for the sake of preventing decoherence.

Once initial cluster state experiments have been performed, it
simply becomes a question of how many cavities one can reasonably
couple together. Alternatively, since the two-qubit gate that we can
generate is entangling (and hence universal for quantum
computation), we can also consider using it directly to implement
the circuit model of computation. This has a much smaller overhead
of qubits, but instead requires much higher quality cavities. For
example, to factor 15 we would only need a $5\times 3$ grid of
cavities to give us six computational qubits. However, we would need
approximately 15 consecutive entangling steps (we have attempted to
minimise this number by allowing as many of the gates to be applied
in parallel as possible, and by optimising the initial labelling of
each qubit), hence requiring a time of order $15\pi/(\sqrt{2}A)$.
Hence, to reduce the effect of dissipative decay, we require an
order of magnitude improvement in the decoherence properties of the
qubits to compensate for the increased running time.

{\em Experimental implementations: }As previously mentioned, there
are three primary candidate technologies; fiber coupled
micro-toroidal cavities \cite{aoki-2006,armani-vahala03}, arrays of
defects in PBGs
\cite{vuckovic-apl,song-noda05, AntonioBadolato05202005} and
superconducting qubits coupled through microwave stripline
resonators \cite{wallraff-2004}. In order to achieve the required
limit of no more than one excitation per site
\cite{angelakis-bose06b}, the ratio between the internal atom-photon
coupling and the hopping of photons down the chain should be of the
order of $g/A\sim10^{2}-10^{1}$($A$ can be tuned while fabricating
the array by adjusting the distance between the cavities and g
depends on the type of the dopant). In addition, the cavity/atomic
frequencies to internal coupling ratio should be $\omega_d,\omega_0
\sim 10^{4}g,10^{5}g$ and the losses should also be small,
$g/\max(\kappa,\gamma)\sim 10^3$, where $\kappa$ and $\gamma$ are
cavity and atom/other qubit decay rates. The polaritonic states
under consideration are essentially unaffected by decay for a time
$10/A$ ($10$ns for the toroidal case and $100$ns for microwave
stripline resonators). While the decay time of $10/A$ may seem uncomfortably close to the
preparation time for a cluster state, $\sqrt{2}\pi/A$, the
previously described technique (Fig.~\ref{panels}) of continuously
reforming the cluster state and connecting it to the output of the
previous stage allows a continuous computation that exceeds the
decay time for an individual cavity. The required parameter values
are currently on the verge of being realised in both toroidal
microcavity systems with atoms and stripline microwave resonators
coupled to superconducting qubits, but further progress is needed.
Arrays of defects in PBGs remain one or two orders of magnitude away, but recent developments, and the integrability of these
devices with optoelectronics, make this technology very promising as
well. In all implementations the cavity systems are well separated
by many times the corresponding wavelength of any local field that
needs to be applied in the system for the measurement process.

 {\em Conclusions: } In this paper, we have shown how universal
quantum computation could be realized in a coupled array of
individually addressable atom-cavity systems, where the qubits are
given by mixed light-matter excitations in each cavity site. While
single-qubit operations can be locally achieved, the only available
interaction between qubits is due to the natural system Hamiltonian.
We show how to manipulate this to give a controlled-phase gate
between pairs of qubits. This allows computation either using the
circuit model, or a measurement-based computation, the latter being
most suited to reducing experimental errors. We have discussed
possible architectures for implementing these ideas using photonic
crystals, toroidal microcavities and superconducting qubits and
point out their feasibility and scalability with current or
near-future technology.

 This work was supported by the
QIP IRC (GR/S821176/01), Clare College Cambridge and the European
Union through the Integrated
 Projects QAP (IST-3-015848), SCALA (CT-015714) and SECOQC.

\end{document}